\documentclass[12pt]{iopart}
\usepackage{graphicx} % Include figure files
\usepackage{iopams}

\begin{document}

\title[Control of multiferroic domains by external electric fields in TbMnO$_3$]{Control of multiferroic domains by external electric fields in TbMnO$_3$}

\author{J. Stein$^1$, M. Baum$^1$\footnote{present address: Fraunhofer INT, Euskirchen, Germany}, S. Holbein$^1$, V. Hutanu$^{2,3}$, A. C. Komarek$^1$, and M. Braden$^1$ }
\address{$^1$II. Physikalisches Institut, Universit\"{a}t zu
K\"{o}ln, Z\"{u}lpicher Stra\ss e 77, D-50937 K\"{o}ln, Germany \\
$^2$ RWTH Aachen University, Institut f\"ur Kristallographie,
D-52056 Aachen, Germany \\
$^3$Forschungszentrum J\"ulich GmbH, J\"ulich Centre for Neutron
Science at MLZ, D-85747 Garching, Germany}

\ead{braden@ph2.uni-koeln.de} \vspace{10pt}
\begin{indented}
\item[]31 March 2015
\end{indented}

\begin{abstract}
The control of multiferroic domains through external electric
fields has been studied by dielectric measurements and by
polarized neutron diffraction on single-crystalline TbMnO$_3$.
Full hysteresis cycles were recorded by varying an external field
of the order of several kV/mm and by recording the chiral magnetic
scattering as well as the charge in a sample capacitor. Both
methods yield comparable coercive fields that increase upon
cooling.

\end{abstract}

\submitto{\JPC}
\maketitle

\section{Introduction: chiral  domains in multiferroic compounds}

Multiferroics continue attracting strong interest due to their
application potential in data storage and processing~\cite{1,2}.
In particular the more recently discovered multiferroic materials,
in which the ferroelectric polarization is directly coupled to the
magnetic order~\cite{2}, are promising, because they allow one to
control the magnetic structure through an external electric field.
In these materials the magnetic structure is rather complex with
non-collinear and frequently chiral arrangement of magnetic
moments. Such a non-collinear arrangement can induce a local
electric dipole through the inverse Dzyaloshinski-Moriya
mechanism, $\vec{P}\propto \vec{r}_{ij}\times(\vec{S}_i
\times\vec{S}_j)$, where $\vec{r}_{ij}$ is the connecting vector
between two neighboring spins $\vec{S}_i$ and $\vec{S}_j$
\cite{3}. For example in a cycloid these local dipoles add up to
macroscopic ferroelectric polarization. This mechanism perfectly
describes the multiferroic phases in many of the transition-metal
oxides multiferroics such as TbMnO$_3$\cite{4}, Ni$_3$V$_2$O$_8$
\cite{5} and MnWO$_4$~\cite{6}. The inverse Dzyaloshinski-Moriya
interaction is not the only possibility to induce ferroelectric
polarization through a canted spin arrangement
\cite{arima,kaplan}. For example for lower symmetry such as in
monoclinic systems polarization may also appear parallel to
$\vec{S}_i \times\vec{S}_j$, which applies to the multiferroic
phases in CuFeO$_2$\cite{arima,cufeo}, CuCrO$_2$\cite{cucro},
RbFe(MoO$_4$)$_2$\cite{rbfmoo} and NaFeSi$_2$O$_6$
\cite{baum-neu}.

In the multiferroic phase there are at least two different
electric domains corresponding to the direction of the
ferroelectric polarization. These electric domains are coupled
with the magnetic structure, which is chiral in most cases. A
cycloid does not exhibit a scalar chirality but a vector
chirality, which can be directly associated with the ferroelectric
polarization. Therefore, the magnetoelectric coupling in the
multiferroic phase allows one to control the chiral magnetism with
the aid of an external electric field. In a future application one
can add an exchange bias coupled ferromagnetic layer on top of the
multiferroic  compound in order to read the multiferroic state
\cite{bibes}. Studying the multiferroic domains in a bulk material
is, however, more difficult and requires a microscopic technique
such as second harmonic generation~\cite{shg1,shg2} or polarized
neutron diffraction~\cite{brown}.

Polarized neutron diffraction is ideally suited to study chiral
magnetism~\cite{brown}. This can be easily understood, because the
neutron is a chiral object itself when its magnetic moment and
momentum are parallel. In general magnetic neutron diffraction
measures the Fourier component of the magnetization distribution,
$\vec{M}$, but only the components perpendicular to the scattering
vector, $\vec Q $, contribute, $\vec{M}_\perp$ with $|\vec
M_\perp|^2= |M_y|^2 + |M_z|^2$. Here we use the common coordinate
system in polarized neutron scattering experiments $-\vec x \, \|
\, \vec Q $, $\vec z$ vertical to the scattering plane, and $\vec
y = \vec z \times \vec x$. The chiral magnetic scattering is
defined as $M_\chi^2 = -i(\vec M_\perp \times \vec M_\perp^*)_x$
as it can be illustrated by considering an ideal helix. For such
ideal helix and $\vec Q $ parallel to the propagation vector
$M_\chi^2 $ is maximal and of the same size as $|\vec M_\perp|^2$.
Using spherical polarization analysis one may analyze the full
polarization matrix, i.e.\ analyzing the neutron polarization in
the outgoing beam parallel to $x$, $y$, or $z$ for any
polarization in the incoming beam. The intensity in such a
channel, $\sigma_{uv}$ with $u$, $v$ = $x$, $y$, or $z$, can be
recorded for up and down polarization, respectively, resulting in
four values in each channel. We denote the down polarization by an
overbar. For the longitudinal cases, $\sigma_{xx}$, $\sigma_{yy}$
and $\sigma_{zz}$, one may distinguish spin-flip (SF) and
non-spin-flip (NSF) processes. The chiral components can be
directly determined in various channels, most easily they are
obtained in the ${xx}$ channels where all the magnetic scattering
contributes to the SF scattering. For these two SF intensities one
finds $\sigma_{x{\bar{x}}}= \vec M_\perp \cdot \vec M_\perp^* -
i(\vec M_\perp \times \vec M_\perp^*)_x$ and $\sigma_{{\bar{x}}x}=
\vec M_\perp \cdot \vec M_\perp^* + i(\vec M_\perp \times \vec
M_\perp^*)_x$. In one case the chiral contribution adds to $|\vec
M_\perp|^2$, while it is subtracted in the other case. In the case
of the ideal helix with the propagation vector parallel to $\vec
Q$ the total intensity will thus appear only in one of the two SF
$xx$ channels. However, this applies only to a mono-domain sample,
while for equal domain distribution the two $\sigma_{x{\bar{x}}}$
and $\sigma_{{\bar{x}}x}$ SF channels become again identical.
Comparing these two SF channels gives thus a direct method to
determine the domain distribution in a chiral magnet.
\cite{brown,baum,finger,yamasaki}.

Concerning the multiferroic materials neutron polarization
analysis was first applied for TbMnO$_3$ to show that one can
control the chiral domains by cooling in an external electric
field~\cite{yamasaki}. Similar temperature dependent measurements
were performed for LiCu$_2$O$_2$~\cite{seki} and MnWO$_4$
\cite{finger,sagayama}. For MnWO$_4$ neutron polarization analysis
was also used to follow multiferroic hysteresis cycles by varying
the external electric field at constant temperature and recording
the chiral domain distribution~\cite{finger,pole,baum2}.

TbMnO$_3$ is the prototypical material for the new class of chiral
multiferroic materials, because it exhibits sizable ferroelectric
polarization and large magnetoelectric coupling combined with a
moderately complex crystal and magnetic structure
\cite{4,got2004,ken2005}. Magnetic order in TbMnO$_3$ sets in at
$\mathrm{T_N}$=42\,K leading to a longitudinal spin-density wave
(SDW) with a propagation vector of $\vec{k}_{inc}$=(0 $\sim$0.28
0) in reduced lattice units of the $Pbnm$ structure. Upon further
cooling below $\mathrm{T_c}$=27.3\,K a second magnetic transition
occurs into an elliptical cycloid phase with nearly the same
propagation vector. In the SDW phase spins are aligned parallel to
the $\boldsymbol{\mathrm{b}}$ direction, while they rotate in the $\boldsymbol{\mathrm{bc}}$ 
plane in the cycloidal phase. This second transition is accompanied by the
onset of ferroelectric order, which is perfectly explained by the
inverse Dzyaloshinski-Moriya mechanism. The same mechanism also
describes the rotation of the ferroelectric polarization in
external magnetic fields~\cite{kim2005}, as the cycloid plane
flops simultaneously from the $\boldsymbol{\mathrm{bc}}$ to the $\boldsymbol{\mathrm{ab}}$ plane
\cite{ali2009}. The multiferroic hysteresis loops have not been
examined in detail for TbMnO$_3$, only a single loop was studied
by second harmonic generation on a thin film~\cite{glavic}.

Here we study the control of the multiferroic domain population
by application of an external electric field in TbMnO$_3$. We
directly measure the ferroelectric polarization by counting the
surface charge in a capacitor and we determine the chiral domain
distribution by polarized neutron diffraction. Both methods
document that it is possible to switch the multiferroic domains in
bulk TbMnO$_3$ by varying the external field at constant
temperature, but only sufficiently close to the ferroelectric
transition.

\begin{figure}[t!]
\includegraphics[width=0.8\columnwidth]{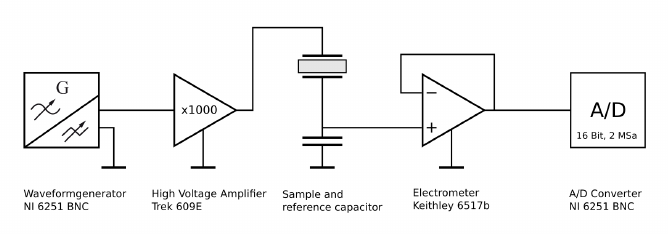}
\caption{Scheme of the setup used for dielectric measurements of
the multiferroic domains in TbMnO$_3$. It consists of a modified
Sawyer-Tower circuit.} \label{Struc0}
\end{figure}

\section{Experimental}

A large single crystal of TbMnO$_3$ was grown with the travelling
solvent floating-zone technique in a mirror furnace (Crystal
Systems Incorporated FZ-T-10000-H-VI-VP). The magnetic transitions
were determined by SQUID measurements on two identically grown
crystals to occur at $T_N$=42.0(2) and $T_C$=27.3(2)\,K in good
agreement with literature~\cite{4,got2004}. All crystals used in
the neutron diffraction and macroscopic studies in this work were
cut from the same single crystal.

Polarized neutron diffraction experiments were performed with the
polarized single-crystal diffractometer POLI-HEIDI installed at
the hot source of the Forschungsreaktor M\"unchen II (FRM-II). The
neutron polarization and analysis is accomplished by $^3$He-spin
filters in the incoming and scattered beams
\cite{hutanu,hutanu1,tasset}. In order to control the decay of the
first filter polarization the incoming beam polarization was
recorded by a transmission monitor. The polarization analysis of
the scattered beam was verified by analyzing a  purely nuclear
Bragg reflection. 
The direction of the neutron polarization at the sample was
controlled with the CRYOPAD setup~\cite{brown}.

\begin{figure}[t!]
\begin{center}
\includegraphics[width=0.8\columnwidth]{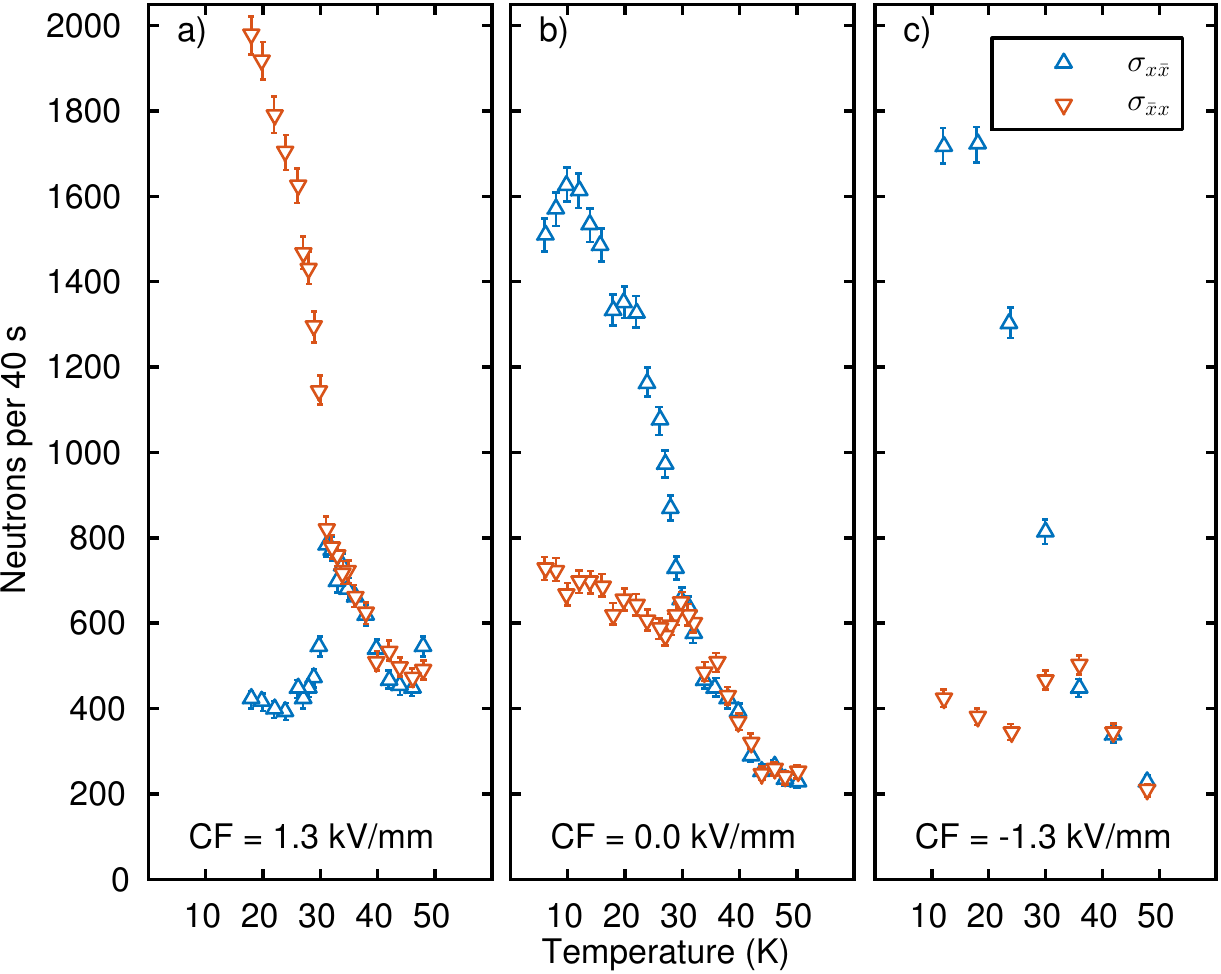}
\end{center} \caption{(Color online) The magnetic scattering at the
(2 0.28 1) Bragg reflection in TbMnO$_3$ was recorded in the two
SF channels for neutron polarization along the $x$ direction.
During cooling through the magnetic transitions, electric fields
of $\pm 1.3$kV/mm were applied in (a) and (c), while the data in
(b) were taken in zero field. The intensities of the two SF
channels $\sigma_{x{\bar{x}}}$ and $\sigma_{{\bar{x}}x}$
interchange upon reversing the field, which indicates the
population of opposite chiral domains.} \label{Struc}
\end{figure}

In order to study the effect of crystal thickness and to apply an
electrical field parallel to the $\boldsymbol{\mathrm{c}}$ direction, two plate-shaped
single-crystals were cut perpendicular to the $\boldsymbol{\mathrm{c}}$ axis with a
thickness of 3.1\ mm and 0.9\ mm, respectively. The electric field
was applied with a capacitor build from aluminum plates and nylon
screws. The sample was mounted in the (2 0 1)/(0 1 0) scattering
plane. The strong magnetic reflection $\vec{Q}$ = (2 0.28 1) was
used for determining the chiral ratio. Note that a large $h$
component of the magnetic Bragg reflection is needed in this
experiment, because both components of the $\boldsymbol{\mathrm{bc}}$ cycloid must
strongly contribute to the scattering, which requires that the $b$
and $c$ axes are nearly perpendicular to $\vec Q$. The intensity
of the thin sample at this reflection was rather low, 50\,cts/s.
The sample was cooled with a standard FRM-II type top-loading
closed-cycle refrigerator, which stabilizes temperatures within
0.1\,K. The challenge of measurements with high electric fields in
croygenic environments consists in fine-tuning the pressure of
helium exchange gas continuously on a thin line where it is possible
to apply several kV/mm without electric discharge while the sample
temperature is still controllable. Much efforts were needed to
determine the breakdown pressure versus voltage curves before
starting the experiment. A temperature offset of about 1\,K was
observed for the phase transitions measured with applied field and
those reported in the literature. This offset arises from the
insufficient thermal contact between the temperature sensor and
the isolating sample, when the exchange gas is lowered to allow
for the application of electric field. In the following we assume
this temperature offset to be constant in the temperature range
from 10 to 50\,K and correct the data correspondingly. The chiral
ratio was determined by measuring the two SF $\sigma_{x{\bar{x}}}$
and $\sigma_{{\bar{x}}x}$ channels as described above.

The macroscopic ferroelectric polarization was measured with a
modified Sawyer-Tower circuit, in which the surface charge is
determined indirectly with an electrometer (Keithley 6517b), see
Figure 1. A thin plate of 0.2\ mm thickness and a contacted
surface area of 9\ mm$^2$ was cut from the same large crystal as
those used in the neutron diffraction experiments. The crystal was
contacted on both sides with conductive silver paint in order to
form the sample capacitor, which is studied in respect to the
reference capacitor in the Sawyer-Tower circuit. We applied
electric fields of up to 3\ kV/mm and recorded the hysteresis
loops in the quasistatic configuration. Running a full loop
corresponds to one second. The high voltage was generated by
symmetric triangular waveforms which have been amplified by a
factor of 1000 in a Trek 609E high voltage amplifier. In a vacuum
of about 10$^{-6}$\,mbar it was possible to apply electric fields
of the order of 5\,kV/mm. The results are limited by finite
conductivity losses, which turned out to be small. The sample
capacitor was mounted on the cold finger of a closed cycle
refrigerator, which allows to reach base temperatures of 5\,K. The
control of temperature and the runs to record the hysteresis loops
were fully automatized. Comparing several hysteresis loops no
ongoing fatigue effects could be detected, but fatigue effects for
an asgrown crystal cannot be excluded. Therefore, it is possible
to average 4 hysteresis loops at each temperature. This procedure
effectively suppresses noise which partially stems from the
triboelectric effect.

\begin{figure}[t!]
\begin{center}
\includegraphics[width=0.5\columnwidth]{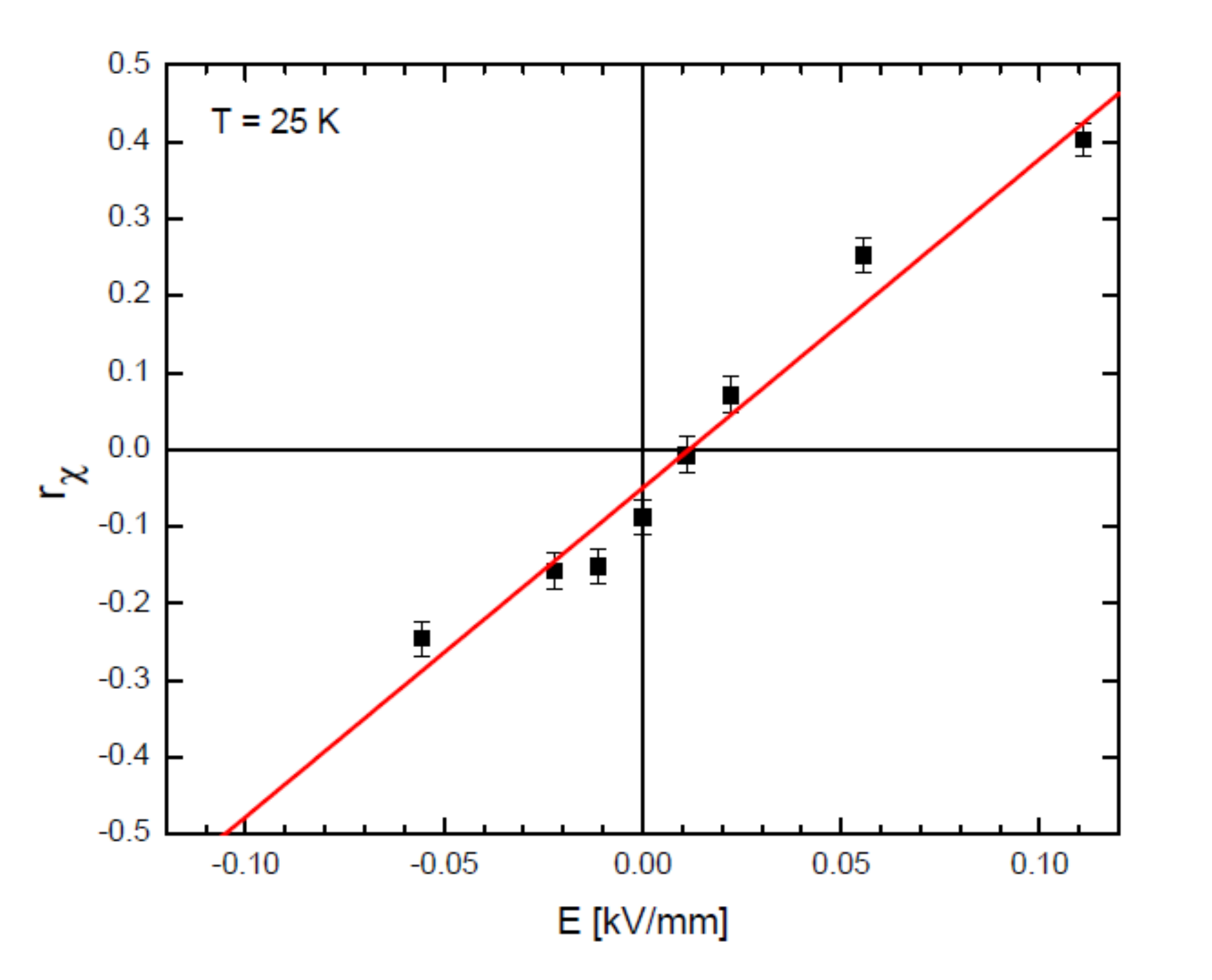}
\end{center}
\caption{(Color online) Chiral ratio measured at T=25\,K after
cooling the  thick sample in small electric fields from the
paramagnetic (50\,K) into the chiral magnetic phase (25\,K). The
sequence was measured with reducing and alternating field
strengths. At low electric fields the chiral ratio determined at
25\,K shows a linear dependence on the electric field applied
during cooling.} \label{Struc1}
\end{figure}

\begin{figure}[t!]
\begin{center}
\includegraphics[width=0.9\columnwidth]{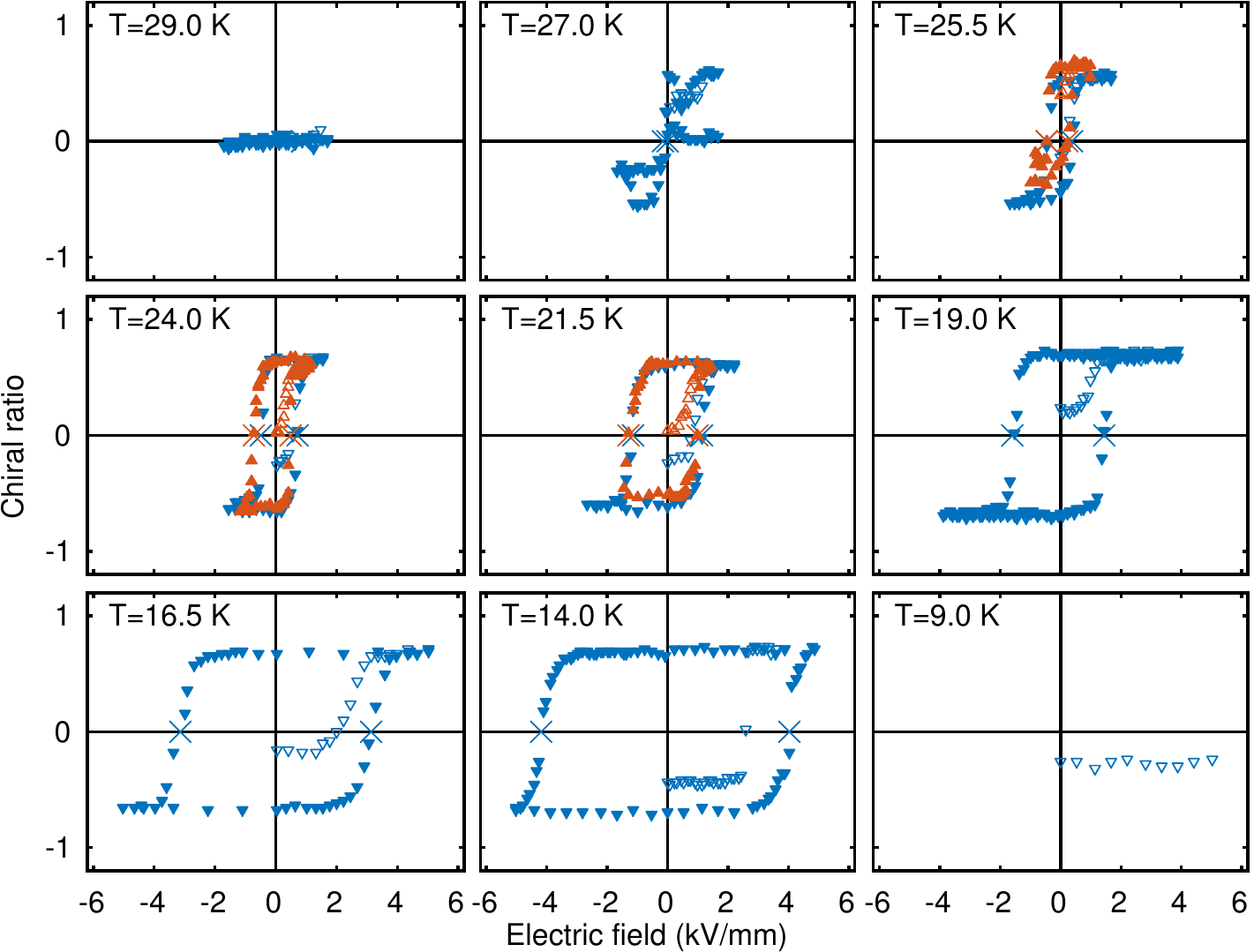}
\caption{(Color online) Hysteresis loops in TbMnO$_3$ obtained by
measuring the chiral ratio as a function of external electric
field at different temperatures for two different samples at
POLI-HEIDI (thicknesses of 0.9 and 3.1\,mm). Open triangles denote
the data taken immediately after cooling and filled symbols denote
the data taken to characterize the full hysteresis loops.
Triangles plotted in orange show the three loops taken with the
thicker crystal. Crosses denote the coercive field values. The
loops were recorded after zero-field cooling from 50\,K.}
\label{Struc2}
\end{center}
\end{figure}

\section{Results}

\subsection{Polarized neutron diffraction experiments}

In a preliminarily test the sample was cooled from 50\,K in
positive and negative electric fields of $\pm$1.3\,kV/mm and in
zero field, see Figure 2. The sample develops a clear preference
for one chiral domain depending on the sign of the electric field.
The intensity of the two spin-flip channels is interchanged by
reversing the electric field. The chiral ratio, $r_{\mathrm{chir}}
=
\frac{\sigma_{x\bar{x}}-\sigma_{\bar{x}x}}{\sigma_{x\bar{x}}+\sigma_{\bar{x}x}}$,
was calculated from the data for cooling in positive field and
roughly follows the development of the electric polarization as it
has been previously reported by the same type of experiment
\cite{yamasaki}. Note that even for a mono-domain state $r_{\mathrm{chir}}$ will deviate from
$\pm$1 due to the finite geometry factors in TbMnO$_3$, which furthermore is not a perfect cycloid
\cite{ken2005}. With the elliptical magnetic structure we
calculate an ideal mono-domain chiral ratio
$r_{\mathrm{chir}}$=$\pm$0.93 for $\vec{Q}$ = (2 0.28 1).

The effect of small electric fields applied during cooling from
the paramagnetic into the chiral magnetic phase is depicted in
Figure 3. The sequence was measured with reducing and alternating
field strengths. Very small fields of a few tenths of V/mm applied
upon cooling are sufficient to produce finite preference for one
chiral domain. However, saturation is not reached for such small
fields. Instead, the chiral ratio shows a linear dependence on
the electric field applied during the cooling, at least in this
low-field range. Note, however, that the chiral response of the
samples is asymmetric with respect to the applied field, as it
even exhibits finite values for zero electric field.

In the following we focus on the possibility to reverse chiral
magnetic domains by varying the electric field at constant
temperature. This represents the multiferroic hysteresis loops,
i.e.\ the control of magnetic order by an external electric field.
The obtained hysteresis loops are shown in Figure 4. All
hysteresis loops were recorded after zero-field cooling from 50\,K
to the respective temperature. To ensure that there was no
electric field applied during cooling the sample surfaces were
short-circuited for most of the loops shown in Figure 4. Only the
curve taken at 19\,K was obtained after cooling the sample with
open contacts, which seems to induce a different direction of
preference. It is a remarkable finding that the initial curves
always start at finite values even for cooling with
short-circuited surface contacts. The hysteresis loops were
recorded by varying the electric field in finite steps and by
recording the neutron intensities at constant fields. Due to the
limited count rates, it was necessary to count several minutes per
point. Therefore, the shown hysteresis loops are quasi-static for
relaxation times of the order of minutes and below.

\begin{figure}[t!]
\begin{center}
\includegraphics[width=0.85\columnwidth]{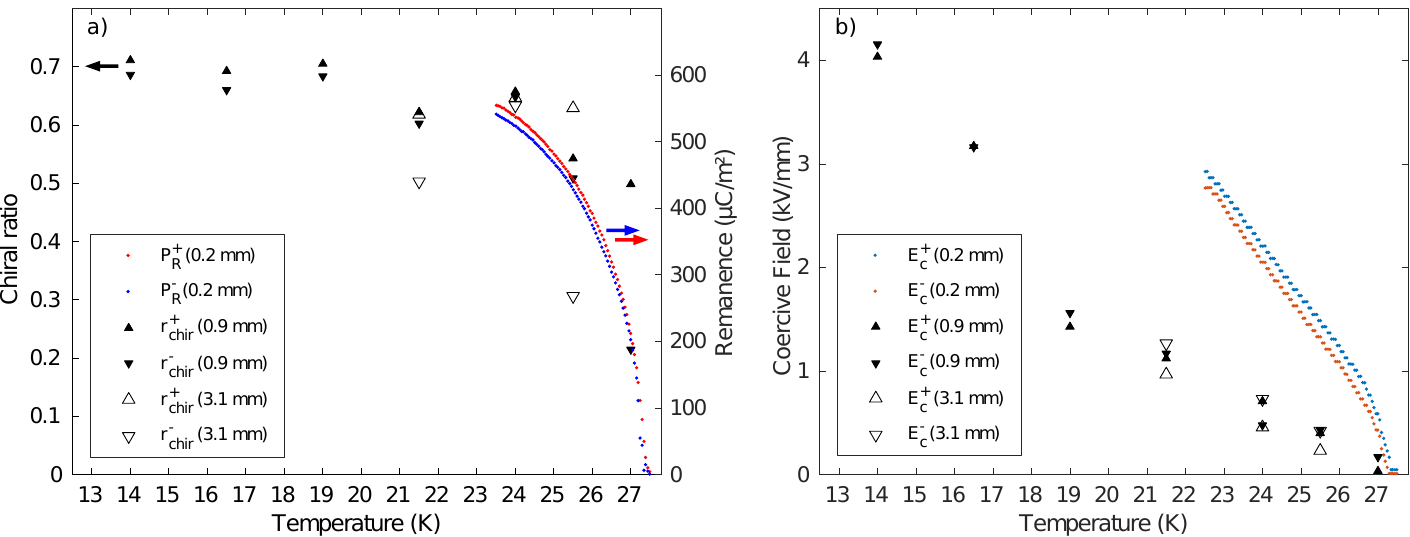}
\end{center}
\caption{(Color online) (a) Temperature dependence of the remanent
ferroelectric polarization in TbMnO$_3$ as obtained from the
dielectric hysteresis loops shown in Fig.~6. In comparison we also
show the temperature dependence of the remanent chiral ratios
observed in polarized neutron diffraction. In both parts larger
triangles correspond to the neutron results, and the tip of the
triangle indicates the direction of the applied electric
fields. (b) Temperature dependence of the coercive fields obtained
in the neutron diffraction and dielectric measurements. The
positive and negative coercive fields obtained from the
chiral-ratio loops match perfectly underlining the symmetric
character of the sample aside from the results obtained for the
thicker crystal with neutron diffraction. Furthermore, the
coercive fields obtained in the neutron experiment do not depend
on sample thickness. The coercive field obtained from the
dielectric experiments are considerably higher.} \label{Struc3}
\end{figure}

Most recorded hysteresis loops show a symmetric shape. Both the
coercive fields and the saturation values are in good accordance
between the two directions. The coercive field increases with
decreasing temperature while an approximately constant saturation
value of $r_{\mathrm{chir}}$ can be reached at low temperatures by
applying a high enough field, see Fig.~5(a) and (b). Note,
however, that the chiral ratio describes the domain population
multiplied with the intrinsic chiral fraction of the magnetic
scattering and, therefore, it is not expected to continuously
increase upon cooling while the ferroelectric polarization follows
$\vec{S}_i\times\vec{S}_j$. The domain dynamics of the sample
becomes stiffer when cooling deeper into the multiferroic phase.
This behavior qualitatively resembles the results reported for
MnWO$_4$~\cite{baum,finger}. At 9\,K the domains are strongly
pinned; therefore, the electric field causes an electrical
breakdown before the chiral ratio can be affected. Already at
14\,K the maximal applicable electric field hardly surpasses the
coercive field needed to start reverting the multiferroic domains
within the time scales of the experiment.

The hysteresis loops of the two samples of different thicknesses
match well each other concerning the coercive fields aside from a
more pronounced asymmetry for the thicker crystal. In order to
apply the same electric field strength to the thicker sample a
larger voltage has to be applied. For that reason it is not
possible to reach as high electric fields for the thicker sample
as for the thinner sample. Therefore, we registered only three
hysteresis loops at higher temperatures for the sample, which was
3.1\,mm thick. The coercive fields for all temperatures are shown
in Figure 5 (b). The positive and negative coercive fields match
well underlining the symmetric character of the thinner samples.
The coercive field shows roughly a linear dependence on the
temperature. There is less agreement concerning the remanent
values of the chiral ratio observed in the two samples of
different thickness. The thicker sample of the neutron experiment
remains less polarized.

\begin{figure}[t!]
\begin{center}
\includegraphics[width=0.9\columnwidth]{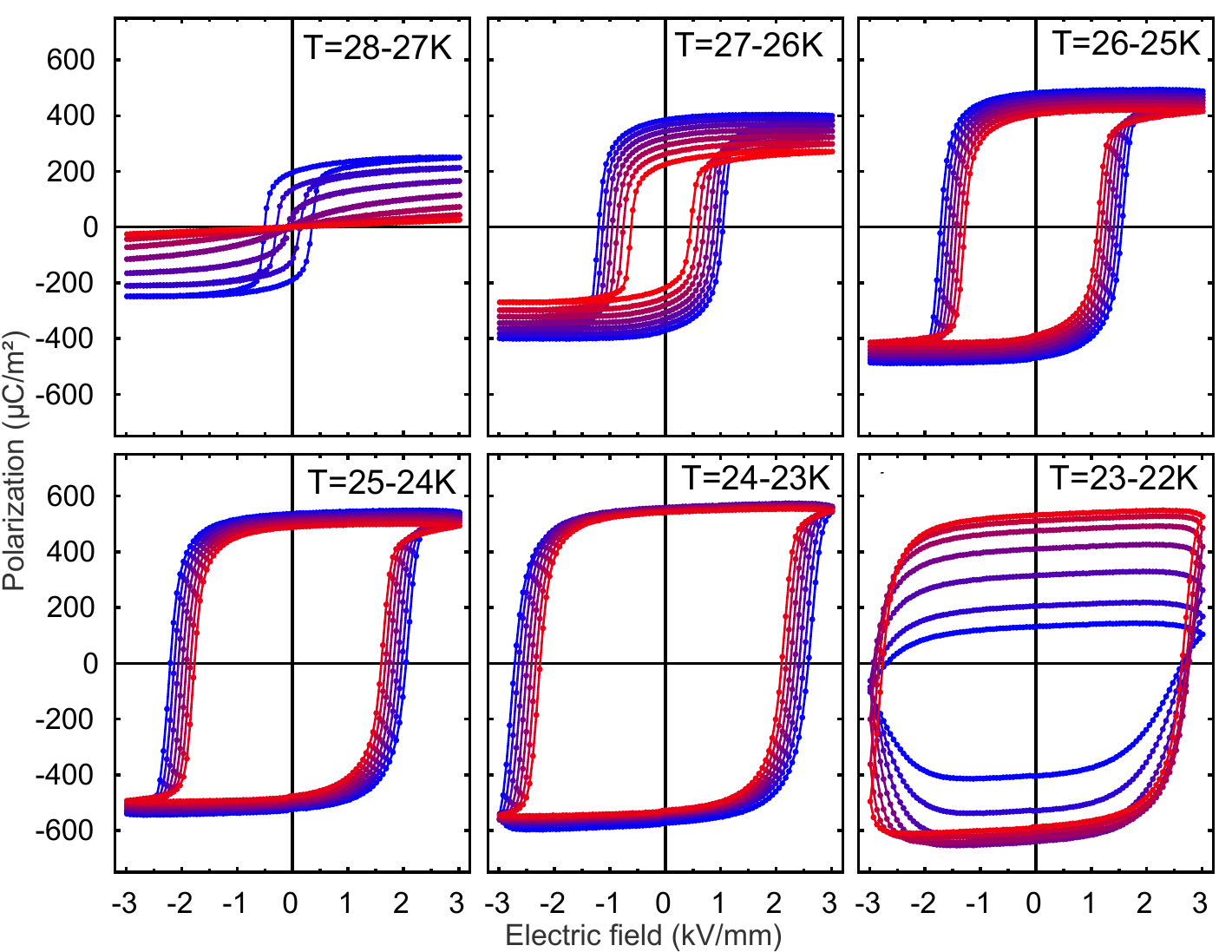}
\end{center}
\caption{(Color online) Dielectric hysteresis loops recorded by
a modified Sawyer-Tower circuit on a 0.2\,mm thick sample. The
colors indicate the temperature varied within each panel in 0.2\,K
steps from the colder (blue) to the warmer (red) temperatures. A
linear background determined above the ferroelectric transition
was substracted.} \label{Struc4}
\end{figure}

\subsection{Ferroelectric hysteresis loops detected by capacitor measurements}

The ferroelectric polarization of TbMnO$_3$ was studied
macroscopically with a modified Sawyer-Tower circuit. The
ferroelectric polarization $P$ versus electric field $E$
hysteresis loops obtained by averaging four successive cycles are
shown in Fig.~6 for different temperatures. The hysteresis loops
are not fully flat in their saturation region due to the
dielectric background contribution, as it is usually observed in
ferroelectrics, and due to finite conductivity effects. Therefore
we have substracted a loop taken in the paraelectric phase from
all the other loops. Finite slopes then still arise from the minor
differences in conductivity and dielectric constant occurring with
the variation of the temperature. For clarity we show only the
data taken in 0.2\,K temperature steps, but intermediate
temperatures were also measured.

The hysteresis loops in Fig.~6 show that the ferroelectric
transition sets in at 27.3(1)\,K in perfect agreement with the
susceptibility data and literature. Below this temperature we find
remanent polarization as well as finite width and area in the
loops. The slight change in the slope of the entire loop can be
attributed to the maximum of the dielectric constant at $T_C$
\cite{4}.

The area of the loop considerably increases upon cooling. The
remanent polarization is almost identical to the saturation
polarization after correcting for the finite slope. The
temperature dependence of the remanent polarization matches the
temperature dependence of the polarization measured by the
pyrocurrent method~\cite{4,got2004} as long as temperatures stay
above 24\,K, see Fig.~5 and 6. This indicates that for
temperatures larger than 24\,K the polarization can be fully
controlled with the maximum applied electric field of 3\,kV/mm and
within a time of about one second. In contrast the data taken
below 24\,K show that saturation polarization cannot be reached
indicating that the relaxation times surpass the times of the
measurement. Between 23 and 24\,K one may recognize that
saturation is not reached before the field starts to decrease from
its maximal values, and below 23\,K the polarization continues to
increase in absolute size even when the electric field is already
decreasing. This clearly shows that the measurement was too fast
to allow the TbMnO$_3$ domains to fully relax. Note that the
neutron experiment due to the low count rates was sufficiently
slow to allow for full relaxation to lower temperatures.

The temperature dependence of the coercive field is shown in Fig.~5(b). 
In this dielectric experiment there is no significant
difference in the coercive fields obtained for the two field
directions, which agrees to the neutron experiments and also to
the symmetric values of the remanent polarization. In the studied
temperature interval the coercive field increases linearly with
temperature after a small jump at the transition. Compared to the
neutron study the coercive fields are considerably larger in spite
of the fact that all the crystals were cut from the same large
single crystal. However, the much thinner sample used in the
dielectric study can more strongly suffer  from the cutting
process. In addition fatigue effects can be stronger in the sample
used for the macroscopic measurement as it went through a much
larger number of thermal and electrical cycles before the shown
experiments were performed. Coercive fields in TbMnO$_3$ are seem
to be sample dependent.

\section{Conclusions}

In conclusion we have shown by polarized neutron diffraction and
by macroscopic measurements that the population of multiferroic
domains in bulk TbMnO$_3$ can be controlled through a moderate
electric field of the order of several kV/mm. Neutron diffraction
determines the reversion of the chiral magnetism while the
macroscopic measurement senses the ferroelectric polarization. The
coercive fields in both studies are of the same order but
significant differences can be ascribed to differences in the
samples used. The chiral magnetic and polarization hysteresis
loops confirm once more the close coupling of the two properties.

\ack

This work was supported through the Bundesministerium f\"ur
Bildung und Forschung through project 05K13PK1 and 05K10PA2. We
are thankful to W. Luberstetter for the technical support during
the neutron diffraction experiment and to S. Masalovich for
supplying $^3$He spin-filter cells.

\end{document}